\documentclass[11pt]{article}
\usepackage[russian,english]{babel}
\usepackage{graphicx,psfrag}
\def\bea{\begin{eqnarray}}
\def\eea{\end{eqnarray}}
\def\nn{\nonumber}
\def\be{\begin{equation}}
\def\ee{\end{equation}}
\def\lb{\label}
\def\bb{\bibitem}

\def\be{\begin{equation}}
\newcommand{\F}{{\cal F}}

\newcommand{\e}{\mbox{\rm e}}

\newcommand{\s}{{\sigma}}
\newcommand{\ka}{{\kappa}}

\begin{document}

\begin{flushright}DTP-MSU/05-11,   LAPTH-1130/05

\end{flushright}

\begin{center}

\vspace{2cm}

{\LARGE {\bf Dyonic  branes  and linear dilaton background} }

\vspace{1cm} {G\'erard Cl\'ement$^{a}$, Dmitri Gal'tsov$^{a,b}$
C\'edric Leygnac$^{a}$ and Dmitri Orlov$^{b}$}

\vspace{1cm} $^{a}${\it Laboratoire de  Physique Th\'eorique LAPTH
(CNRS),
\\
B.P.110, F-74941 Annecy-le-Vieux cedex, France}\\
$^{b}${\it Department of Theoretical Physics, Moscow State
University, 119899, Moscow, Russia}

\vspace{0.5cm}  e-mails: {gclement@lapp.in2p3.fr; gdmv04@mail.ru;
leygnac@lapp.in2p3.fr; orlov{\_}d@mail.ru}

\end{center}

\vspace{1cm}

\begin{abstract}
We study dyonic solutions to the gravity-dilaton-antisymmetric
form equations with the goal of identifying new $p$-brane
solutions on the fluxed linear dilaton background. Starting with
the generic solutions constructed by reducing the system to
decoupled Liouville equations for certain values of parameters, we
identify the most general solution whose singularities are hidden
behind a regular event horizon, and then explore the admissible
asymptotic behaviors. In addition to known asymptotically flat
dyonic branes, we find two classes of asymptotically non-flat
solutions which can be interpreted as describing magnetically
charged branes on the electrically charged linear dilaton
background (and the $S$-dual configuration of electrically charged
branes on the magnetically charged background), and uncharged
black branes on the dyonically charged linear dilaton background.
This interpretation is shown to be consistent with the first law
of thermodynamics for the new solutions.

\end{abstract}
\vfill \eject

\section{Introduction}
Holographic dualities which relate  classical supergravities with
quantum field theories in lower dimensions were first discovered
in non-dilatonic theories (AdS/CFT correspondence) \cite{adscft}
and then extended to the generic case of theories with a dilaton
\cite{ItMaSoYa98}. An important step in establishing these
dualities  consists in the study of the near-horizon limit of
$p$-branes. In the case of dilatonic branes the near-horizon
geometry is either anti-de Sitter (AdS) or Minkowski with a
non-trivial dilaton field depending linearly on an appropriate
radial coordinate which we call in what follows the linear dilaton
background (LDB). Such configurations are 1/2 supersymmetric in
the context of supergravities (contrary to maximally
supersymmetric ones in the case of non-dilatonic branes), but the
conformal symmetry is broken by the dilaton. These backgrounds are
dual to non-conformal QFT-s with sixteen supercharges living on
their boundary \cite{ItMaSoYa98}. In the particular case of the
NS5 brane the corresponding dual theory is the  little string
theory \cite{Ah99}, while in the general case one finds a class of
theories exhibiting the Domain-Wall/QFT correspondence
\cite{Boonstra:1998mp}. By the standard argument, the {\it
thermal} version of the dual quantum theory should have as a
holographic dual the linear dilaton background endowed with  an
event horizon. A variety of relevant supergravity configurations
were obtained both in the black hole case
\cite{ChHoMa94,ClGaLe02,ClLe04,Leygnac:2004bb} and for general
$p$-branes \cite{Clement:2004ii,Gal'tsov:2005vf,ChGaOh05}. These
solutions have a non-trivial electric or magnetic field which is
attributed to the LDB background, while the presence of the event
horizon is interpreted as due to a neutral $p$-brane on this
background.

A natural question arises whether there exist {\it charged} branes
on the linear dilaton background. To answer this, here we study
systematically the {\em dyonic} brane configurations supported by
a unique form field with both electric and magnetic sectors
non-empty. It is worth noting that branes with both electric and
magnetic charges may exist in any space-time with electric and
magnetic branes having different dimensions (branes within branes
of the type of Ref. \cite{Co96}. In even dimensions and with an
antisymmetric form of a suitable rank, electric and magnetic
branes may both have the same dimension \cite{Izquierdo:1995ms}.
Here we will be interested by dyonic branes of this latter type,
which are derivable from Liouville systems. Some dyonic branes can
also be identified within the class of intersecting branes for
suitable values of the parameters \cite{GI,IvMe01,ohta}, though in
this approach the Chern-Simons terms in the Bianchi identities
(transgressions), which can also be relevant for dyonic branes,
are usually not taken into account (for a more detailed discussion
see \cite{Izquierdo:1995ms,Cv01}). We will restrict ourselves here
to dyonic configurations which exist in even space-time dimensions
$d=2n$ in presence of a form field of rank $q=p+2=n$.
Asymptotically flat (AF) branes possessing both electric and
magnetic charges were discussed in a number of papers
\cite{DuLuPo96,LuPoXu96,Lu:1996gt,Lu:1995yn}, some non-AF
solutions were also mentioned in
\cite{Grojean:2001pv,Yazadjiev:2005du}.

Our strategy consists in obtaining the generic solution of the
supergravity field equations for $ISO(p)$ symmetric branes with
the transverse space being the product of a homogeneous space of
dimension $k$ and a flat $(q-k)$- dimensional Euclidean space. Such
a possibility exists for certain particular values of the dilaton
coupling constant, when the system of equations can be reduced to
decoupled Liouville equations. We then demand the absence of
naked singularities without imposing any asymptotic conditions.
The resulting solution possesses an (either non-degenerate or
degenerate) event horizon and can be interpreted as a black
brane. Then we explore all possible asymptotic behaviors
in the region at an infinite geodesic distance from the horizon
and find three different classes. The first class consists of the
usual asymptotically flat black branes possessing both electric
and magnetic charges. The solutions of the second class are
magnetically charged black branes on the electrically charged LDB
(and the dual electrically charged black branes on the
magnetically charged LDB). The last class contains uncharged
branes on the dyonic LDB.

In order to test this interpretation we develop the thermodynamics
of our general dyonic configurations. To calculate the brane
tension and other physical characteristics of the asymptotically
non-flat solutions one needs to generalize the formalism of
quasilocal charges developed in particular in Refs.
\cite{Brown:1992br,HaHo95,ChNe98} to the case of an arbitrary number of
space-time dimensions and to the presence of the antisymmetric
form fields. This was done in Ref. \cite{Clement:2004ii}, so we can
directly apply this technique here.  We find that the
asymptotically flat dyons satisfy the expected first law including
variations of both the electric and magnetic charges. For the
second class the first law includes only the variation of the
magnetic charge, while the electric field remains frozen. This
fits nicely with the expectation that we deal with a magnetic
brane on the electric LDB (similarly, with an electric brane on
the magnetic LDB). Finally, for the third class of solutions,
neither electric , nor magnetic charge variations contribute to
the first law. Hence both these charges must be attributed indeed
to the background, not to the brane.

\section{Setup}

\setcounter{equation}{0} Our starting point is the standard action
for the metric, dilaton and an antisymmetric form
\bea\label{action} S = \int d^d x \sqrt{-g} \left( R - \frac12
\partial_\mu \phi
\partial^\mu \phi - \frac1{2\, q!} \, {\rm e}^{a\phi} \, F_{[q]}^2
\right)\,,
\eea
with the Newton constant $G = 1/16\pi$.

We consider $p$-brane spacetime configurations with a $p+1$-dimensional
world volume and the $q$-dimensional transverse space
$\Sigma_{k,\sigma}\times R^{q-k}$ ($p+q=d-2$):
\bea
\label{metric}
ds^2 = - {\rm e}^{2 B} dt^2 + {\rm e}^{2 D} (dx_1^2 + \cdots +
dx_p^2) +{\rm e}^{2 A} dr^2+ {\rm e}^{2 C} \, d\Sigma_{k,\sigma}^2 +
{\rm e}^{2 E}(dy_1^2 + \cdots +dy_{q-k}^2),
\eea
where the metric functions $A, \cdots, E$ depend only on the radial
coordinate $r$, and $\Sigma_{k,\sigma}$ is a constant curvature space ($k
> 1$):
\be
d\Sigma_{k,\sigma}^2 = \bar g_{ab} dz^a dz^b = \left\{
 \begin{array}{ll}
d \varphi^2 + \sin^2\varphi \, d\Omega_{(k-1)}^2, \qquad & \sigma=+1, \\
d \varphi^2 + \varphi^2 \, d\Omega_{(k-1)}^2, \qquad & \sigma=0,\\
d \varphi^2 + \sinh^2\varphi \, d\Omega_{(k-1)}^2, \qquad & \sigma=-1.
 \end{array} \right.
\label{gmetric}
\ee
The Ricci tensor for the metric
(\ref{metric}) has the non-vanishing components
\bea
R_{tt} &=& e^{2B-2A}\left[B''+B'(-A'+B'+kC'+pD'+(q-k)E')\right], \\
R_{\alpha\beta} &=& -e^{2D-2A}\left[ D'' +
D'(-A'+B'+kC'+pD'+(q-k)E')\right]\delta_{\alpha\beta}, \\
R_{rr} &=& -B''-B'(B'-A')-k(C''+C'^2-A'C')\nn\\ && -p(D''+D'^2-A'D')
-(q-k)(E''+E'^2-A'E'), \\
R_{ab} &=& -\left\{e^{2C-2A}\left[C''+C'(-A'+B'+kC'+pD'+(q-k)E')\right]
-\sigma(k-1)\right\}\,\bar g_{ab}, \\
R_{ij} &=& -e^{2E-2A}\left[E''+
E'(-A'+B'+kC'+pD'+(q-k)E')\right]\delta_{ij},
\eea
Dyonic configurations for the $q$-form field $F_{[q]}$ are possible in
an even dimensional spacetime $d=2n$, with
\be
q = p+2 = n.
\ee
In this case the Maxwell equations and the Bianchi identities are solved by
\bea
F_{[n]}=b_1\mbox{\rm vol}_n+b_2\,\e^{-a\phi}*\mbox{\rm vol}_n,\quad\mbox{\rm
vol}_n={\rm vol}(\Sigma_{k,\sigma})\wedge dy_1 \wedge \cdots \wedge dy_{n-k}.
\eea
Define the  gauge function $\F$
\be \lb{gauge} \ln
\F \equiv -A+B+kC+pD+(q-k)E. \ee
Fixing the form of $\F$ we thereby choose some gauge condition. The field
equations are particularly simple to solve in the gauge $\F = 1$. The
corresponding radial coordinate $\rho$ is related to the radial
coordinate $r$ in a generic gauge $\F$ by
\be\lb{transF}
dr = \F d\rho.
\ee
Denoting the derivatives with respect to $\rho$ by a dot, and putting
\be\lb{GiH}
G_1=a\phi+2B+2(n-2)D,\quad G_2=-a\phi+2B+2(n-2)D, \quad H = 2(A-C),
\ee
we obtain the following form for the Einstein equations and the dilaton field
equation
\bea
&&\ddot{B}=\frac14\left\{b_1^2\e^{G_1}+b_2^2\e^{G_2}\right\}
\label{eqBd}\\
&&\ddot{D}=\frac14\left\{b_1^2\e^{G_1}+b_2^2\e^{G_2}\right\}  \\
&&\ddot{A}=-\frac14\left\{b_1^2\e^{G_1}+b_2^2\e^{G_2}\right\}+
\sigma k(k-1)\e^{2H} \label{eqAd}\\
&&\ddot{C}=-\frac14\left\{b_1^2\e^{G_1}+b_2^2\e^{G_2}\right\}+
\sigma(k-1)\e^{2H} \\
&&\ddot{E}=-\frac14\left\{b_1^2\e^{G_1}+b_2^2\e^{G_2}\right\}  \\
&&\ddot{\phi}=\frac{a}{2}\left\{b_1^2\e^{G_1}-b_2^2\e^{G_2}\right\},
\eea
together with the constraint
\be\label{ceqd}
-\dot{A}^2+\dot{B}^2+k\dot{C}^2+(n-2)\dot{D}^2+
(n-k)\dot{E}^2+\frac12\dot{\phi}^2
=\frac{b_1^2}2\e^{G_1}+
\frac{b_2^2}2\e^{G_2}-\sigma k(k-1)\e^{2H}.
\ee

From the above system, we obtain the equations for the functions $G_1$,
$G_2$  and $H$:
\bea
\ddot{G_1}&=&\frac1{2}\left\{b_1^2\Delta_1\e^{G_1}+
b_2^2\Delta_2\e^{G_2}\right\} \lb{G1}\\
\ddot{G_2}&=&\frac1{2}\left\{b_1^2\Delta_2\e^{G_1}+
b_2^2\Delta_1\e^{G_2}\right\} \lb{G2}\\
\ddot{H}&=&2\sigma(k-1)^2 \e^{H}, \lb{H}
\eea
where
\be
\Delta_1=a^2+(n-1), \quad \Delta_2=-a^2+(n-1).
\ee

The first two equations decouple in three special cases
\cite{DuLuPo96}. The obvious first case is $a^2=n-1$ ($\Delta_2 =
0$). The other two possibilities correspond both to \be G_1-G_2
\equiv 2a\phi = 2a\phi_0 \ee constant. Substracting (\ref{G2})
from (\ref{G1}), we find that this is possible if \be 0 =
a^2\left(b_1^2\e^{a\phi_0}-b_2^2\e^{-a\phi_0}\right)\e^{\frac{G_1+G_2}2}\,,
\ee that is, if either $a$ arbitrary with $b_1^2\e^{a\phi_0} =
b_2^2\e^{-a\phi_0}=b_1b_2$ (we assume without loss of generality
$b_1b_2 > 0$), or $a=0$ ($\Delta_1=\Delta_2$). In the first case
($a^2=n-1$) the equations for $G_1$ and $G_2$ separate and give
\be \ddot{G}_{1,2}=b_{1,2}^2(n-1)\e^{G_{1,2}}\,, \ee so the
solution will be \be\lb{solG1}
G_{1,2}=\ln\left[\frac{\alpha_{1,2}^2}{2(n-1)b_{1,2}^2}\right]
-\ln\left[\sinh^2
\left(\frac{\alpha_{1,2}}2(\rho-\rho_{1,2})\right)\right], \ee
with integration constants $\alpha_{1,2}$ (real or imaginary), and
$\rho_{1,2}$. In the second case ($a\neq0$) and the third case
($a=0$), $G_{1,2}=G\pm a\phi_0$, with $G$ obeying the equation \be
\ddot{G}=b^2(n-1)\e^G\,, \ee where $b^2=b_1b_2$ in the second
case, and $b^2=(b_1^2+b_2^2)/2$ in the third case. The solution
can be written in the form (\ref{solG1}) with $\rho_1=\rho_2$,
$\alpha_1=\alpha_2$, and $b_{1,2}$ replaced by $b^2=b_1b_2$ in the
second case and by $b=\sqrt{(b_1^2+b_2^2)/2}$ in the third case.
The second case is a subset of the third for $b_1=b_2$ which can
be recovered  by a dilaton shift and a redefinition of $b_{1,2}$
as follows: \bea \phi\to \phi-\phi_0,\qquad b_{1,2}\to b_{1,2}e^{\pm
a\phi_0}. \eea In all cases, the solution of the equation
(\ref{H}) is \bea\lb{solH} H=\left\{
\begin{array}{ll}
\ln[\frac{\beta^2}{4(k-1)^2}]-\ln[\sinh^2(\frac{\beta}2\rho)],
&  \quad \sigma=1, \\
\beta\rho,
&  \quad \sigma=0, \\
\ln[\frac{\beta^2}{4(k-1)^2}]-\ln[\cosh^2(\frac{\beta}2\rho)],
&  \quad \sigma=-1,
\end{array}\right.
\eea with an integration constant $\beta$ (real or imaginary for
$\s = +1$, real for $\s= 0$ or $-1$); we have used the translation
freedom inherent in the definition (\ref{transF}) of $\rho$ to set
the second integration constant to zero. In the limiting cases
$\alpha_{1,2} = 0$ and $\beta=0$ the solutions (\ref{solG1}) and
(\ref{solH}) should be replaced by: \bea
G_{1,2}&=&-\ln[(n-1)b_{1,2}^2(\rho-\rho_{1,2})^2/2], \\
H&=&\left\{
\begin{array}{cl}
-\ln(\rho^2), & \quad \sigma=1, \\
H_0, &  \quad \sigma=0,
\end{array}\right.
\eea
The final solution is given in terms of $G_1$, $G_2$ and $H$ as
\bea
B&=&\frac1{4(n-1)}\{G_1+G_2\}+(n-2)\{d_1\rho+d_0\}, \lb{fdB}\\
D&=&\frac1{4(n-1)}\{G_1+G_2\}-\{d_1\rho+d_0\}, \lb{fdD}\\
A&=&-\frac1{4(n-1)}\{G_1+G_2\}+\frac{k}{2(k-1)}H-\frac{n-k}{k-1}
(c_1\rho+c_0), \lb{fdA}\\
C&=&-\frac1{4(n-1)}\{G_1+G_2\}+\frac1{2(k-1)}H-\frac{n-k}{k-1}
(c_1\rho+c_0), \lb{fdC}\\
E&=&-\frac1{4(n-1)}\{G_1+G_2\}+c_1\rho+c_0,\lb{fdE}\\
a\phi & = & \frac12\{G_1-G_2\}.\lb{fdf}
\eea
The integration constants are related by the constraint equation
\be\lb{cons}
\frac1{4(n-1)}(\alpha_1^2+\alpha_2^2)-\frac{k}{4(k-1)}\beta^2
+\frac{(n-1)(n-k)}{k-1}c_1^2 +(n-1)(n-2)d_1^2 = 0\,.
\ee
Furthermore, we can always rescale the $x$ and $y$ coordinates so that
\be
c_0=d_0=0.
\ee

\section{Black dyons}
\setcounter{equation}{0}

From now on we will consider the case of the spherical topology of
the transverse space, $\sigma=1$. Generically, the space-time will
contain an event horizon which can be identified with the surface
$\rho=-\infty$. On the horizon, choosing the affine parameter
$\lambda$ along the radial geodesic as \bea
d\lambda=\e^{A+B}d\rho, \eea one must have for the metric function
$B$ \bea\lb{affhor} \e^{2B}\sim\lambda^m, \eea where $m=1$ for a
non-degenerate horizon, and $m=2$ in the degenerate case. Assuming
first $\alpha_{1,2}$ and $\beta$ real positive, we find near the
horizon \bea G_{1,2}\approx \alpha_{1,2}\rho+\mbox{\rm const.}\,,
\quad H\approx \beta\rho+\mbox{\rm const.}, \eea Therefore,
$\e^{2B}$ vanishes on the horizon provided \bea \alpha_1+\alpha_2
+ 4(n-1)(n-2)d_1 > 0. \eea

\paragraph{ Non-degenerate case.}
Differentiating (\ref{affhor}), one obtains
\be
\frac{d\e^{2B}}{d\lambda}=2\e^{B-A}\dot{B} \sim 1,
\ee
and hence $B-A \to$ const. It follows that the left-hand side of the
constraint equation (\ref{ceqd}) reduces on the horizon to a sum of
squares, while the right hand side goes to zero, so equating the separate
terms to zero one obtains
\bea\lb{alphabet}
\alpha_1=\alpha_2=\beta=2(n-1)c_1=2(n-1)d_1\,.
\eea

The black solution can be transformed to a Schwarzschild-like gauge by
the map \bea \lb{tauxi} \e^{\alpha\rho} = \frac{f_+(\xi)}{f_-(\xi)},
\eea with \be f_{\pm}(\xi) = 1 - \frac{\xi_{\pm}}{\xi}, \ee so that
the horizon $\rho \to -\infty$ maps to a finite value $\xi =
\xi_+$. This map defines $\xi$ (and its special values $\xi_{\pm}$)
only up to a scale, which we shall fix such that \bea \lb{axipm}
\xi_+-\xi_- = \frac{\alpha}{k-1}.  \eea The images of $\rho_1$ and
$\rho_2$ under the map (\ref{tauxi}) define the new integration
constants $\xi_1$ and $\xi_2$, \bea \lb{rho12}
\e^{\alpha\rho_{1,2}}=\frac{f_+(\xi_{1,2})}{f_-(\xi_{1,2})}, \eea
which (due to the positivity of the left-hand side) lie both outside
the interval $[\xi_-,\xi_+]$, i.e.  \be\lb{rangexi12} \xi_{1,2}
\subset (-\infty,\xi_-) \cup (\xi_+,+\infty). \ee

It is convenient to further transform the radial coordinate to $r$,
with \bea \lb{xir} \xi=r^{k-1}. \eea This corresponds to fixing the
following gauge function \bea\lb{garhor} \F=r^k f_+f_-. \eea The
solution then takes the form \bea
ds^2&=&\left[\frac{\e^{G_0}f_{-}^2}{f_{1}f_{2}}\right]^{\frac1{n-1}}
\left\{-\frac{f_{+}}{f_{-}}dt^2+d{\bf x}^2\right\} \nn\\
&+&\left[\frac{\e^{G_0}f_{-}^{2}}{f_{1}f_{2}}\right]^
{-\frac1{n-1}}\bigg[(f_-^2)^{\frac1{k-1}}
\left\{\frac{dr^2}{f_{+}f_{-}}+r^2d\Omega^2_{k}\right\}+ d{\bf
y}^2\bigg], \lb{bdm} \\
e^{a\phi}&=&e^{a\phi_0}\frac{f_{2}}{f_{1}},\lb{bdphi}\\
F_{[n]}&=&b_1\mbox{vol}(\Omega_{k})\wedge dy_1\wedge\cdots\wedge
dy_{n-k} \nn \\ &&\qquad\qquad-b_2\e^{G_0-a\phi_0}\,\frac{dr}
{f_2^2r^k}\wedge dt\wedge dx_1\wedge\cdots\wedge dx_{n-2}\,, \lb{bdF}
\eea with \be f_{1,2} = 1-\frac{\xi_{1,2}}{\xi}, \ee and \bea \e^{G_0}
&=&\frac{2(k-1)^2}{(n-1)b_1b_2}
\bigg[(\xi_{+}-\xi_{1})(\xi_{-}-\xi_{1})(\xi_{+}-\xi_{2})(\xi_{-}-\xi_{2})
\bigg]^{1/2},\lb{eG0}\\
e^{a\phi_0}&=&\frac{b_2}{b_1}\bigg[\frac{(\xi_{+}-\xi_{1})(\xi_{-}-\xi_{1})}
{(\xi_{+}-\xi_{2})(\xi_{-}-\xi_{2})}\bigg]^{1/2}\lb{eaphi0}\,.  \eea

The values $\xi = \xi_{1,2}$ correspond to curvature
singularities. This follows from the fact that the Ricci scalar, which
from the Einstein equations for our dyons is simply \bea
R=\e^{-2{A}}\frac{\dot{\phi}^2}{2}, \eea behaves as  \bea
R\sim[f_1f_2]^{-\frac{2n-1}{n-1}} \eea for $a^2=n-1$, while the
Kretschmann scalar (in both cases $a=0$ and $a^2=n-1$) behaves as \bea
R_{\alpha\beta\gamma\delta}R^{\alpha\beta\gamma\delta}\sim
[f_1f_2]^{-2\frac{2n-1}{n-1}}.  \eea  So, for finite $\xi_{1,2}$
($\rho_{1,2}\neq 0$), the black solution will be regular only if both
$\xi_1$ and $\xi_2$ lie behind the outer horizon $\xi_+$, which (owing
to (\ref{rangexi12})) implies \be \xi_{1} \le \xi_2 < \xi_- < \xi_+
\ee (the permutation of $\xi_1$ and $\xi_2$ will lead only to the sign
change of the dilaton $\phi \to-\phi$).

\paragraph{Degenerate horizon.}
In terms of the affine parameter, the metric function  $\e^{2B}$ and
its derivative behave near the horizon as \bea
\e^{2B}\sim\lambda^2,\e^{B-A}\dot{B}\sim\lambda, \eea so one obtains:
\bea \e^{-A}\dot{B}\sim O(1). \eea Two possibilities then exist:
either both terms are regular, or $\dot{B}$ and $\e^A$ both vanish on
the horizon.  Assuming the regularity of $\e^A$, one finds from
(\ref{fdA}) that on the horizon $\dot{A}=0$, and the same for
$\dot{B}$ from the constraint equation, which contradicts the
assumption, so we have the second case.

If both $\dot{B}=0$ and $\e^{B}=0$, then at least one of the
parameters $\alpha_{1,2}$ must vanish. Then $\e^A=0$ on the horizon is
possible only if $\beta=0$. It then follows from the constraint
equation (\ref{cons}) that the remaining parameters vanish. So the
degenerate solution corresponds to the conditions \bea
\alpha_1=\alpha_2=\beta=c_1=d_1=\varphi_1=0. \nn \eea One can check
that this is equivalent to taking the limit $\xi_{-}\to\xi_+$ in the
previous solution.

\section{Asymptotic behavior: the three classes of black dyons}
\setcounter{equation}{0}

We have not yet discussed the asymptotic behavior of our solutions at
spatial infinity. Inspection of (\ref{bdm}) shows that spatial infinity
corresponds to  $\xi\rightarrow+\infty$ ($\rho \to 0$), and that the
solution is asymptotically flat. Note however that the general black
dyonic solution (\ref{bdm})-(\ref{bdF}) was written down for non-zero
values of the integration constants $\rho_{1,2}$, according to
(\ref{rho12}). In the special cases in which one or both of these
integration constants vanish, the solution, which is no longer
asymptotically flat, can be recovered from (\ref{bdm})-(\ref{bdF}) by
taking the limit in which one or both of the image constants $\xi_{1,2}$ is
sent to $-\infty$ \cite{Gal'tsov:2005vf}. In the following, we discuss
the three classes of solutions (1): $\rho_1$ and $\rho_2$ are both non
zero; (2): $\rho_1=0$, $\rho_2 \neq 0$ (this can occur only in the
case $a^2 = n-1$); (3) $\rho_1=\rho_2=0$.

\paragraph{First class: asymptotically flat dyons.}
We first consider the generic case with two non-vanishing parameters
$\rho_1$ and $\rho_2$. The metric (\ref{bdm}) is asymptotically Minkowskian
provided
\be\lb{G00}
G_0 = 0.
\ee
Also, the map (\ref{tauxi}) defines $\xi$ only up to an additive constant,
which we can choose so that $\xi_2=0$. Furthermore, the value of the
dilaton at infinity $\phi_0$ can be set to zero by the dilaton shift,
together with the form rescaling
\be\lb{shift0}
\phi \to \phi - \phi_0, \quad F_{[n]} \to \e^{a\phi_0/2}F_{[n]},
\ee
leading to the form of the asymptotically flat solution, for $a^2 = n-1$,
\bea
ds^2&=&\left[\frac{f_{-}^2}{f_{1}}\right]^{\frac1{n-1}}
\left\{-\frac{f_{+}}{f_{-}}dt^2+d{\bf x}^2\right\} \nn\\
&+&\left[\frac{f_{-}^{2}}{f_{1}}\right]^
{-\frac1{n-1}}\bigg[(f_-^2)^{\frac1{k-1}}
\left\{\frac{dr^2}{f_{+}f_{-}}+r^2d\Omega^2_{k}\right\}+ d{\bf y}^2\bigg],
\lb{bdm1}
\\ e^{a\phi}&=&\frac{1}{f_{1}},\lb{bdphi1}\\
F_{[n]}&=&\sqrt{\frac2{n-1}}(k-1)\bigg[\sqrt{(\xi_+-\xi_1)(\xi_--\xi_1)}
\mbox{vol}(\Omega_{k})\wedge dy_1\wedge\cdots\wedge dy_{n-k} \nn \\
&&\qquad\qquad-\sqrt{\xi_+\xi_-}\,\frac{dr}{r^k}\wedge dt\wedge
dx_1\wedge\cdots\wedge dx_{n-2}\bigg]\,,\lb{bdF1}\,,
\eea
depending on the three independent parameters $\xi_+$, $\xi_-$ and $\xi_1$.
Note that the original parameters $b_1$ and $b_2$ have been eliminated
altogether from the solution. The ``magnetic" and ``electric" charges
associated with the solution (\ref{bdm1})-(\ref{bdF1}) are
\bea
&& P = L_pL_{q-k}\Omega_k
\sqrt{\frac2{n-1}}(k-1)\sqrt{(\xi_+-\xi_1)(\xi_--\xi_1)}\,,\\
&& Q = L_pL_{q-k}\Omega_k
\sqrt{\frac2{n-1}}(k-1)\sqrt{\xi_+\xi_-}\,,
\eea
where $L_p$ and $L_{q-k}$ are the normalization volumes of the spaces
spanned by  $x$ and $y$, and $\Omega_k$ is the volume of the unit sphere.
The Ricci scalar for this spacetime is
\be
R =
\frac{(k-1)^2\xi_1^2}{2(n-1)}\,\xi^{-2-\frac1{n-1}}(\xi-\xi_1)^{-2-\frac1{n-1}}
(\xi-\xi_-)^{\frac{nk-3n+k+1}{(n-1)(k-1)}}(\xi-\xi_+)\,.
\ee
This diverges on the inner horizon $\xi = \xi_-$ only for $k=2$ and
$n>3$. However consideration of the Kretschmann scalar shows that the
inner horizon is regular only for $k = n = 2$ or 3.
In these cases the timelike singularity is located at $\xi =
\mbox{\rm sup}(\xi_1,0)$. In all other cases $\xi = \xi_-$ is a
spacelike singularity.

In the second and third cases ($\phi=\phi_0$), $\rho_1=\rho_2$ implies
that also $\xi_1=0$, so that $f_1=1$ in (\ref{bdm1}) and (\ref{bdphi1}). The
only difference between these two cases lies in the number of
independent parameters. In the second case ($a$ arbitrary), the
solution given by (\ref{bdm1})-(\ref{bdF1}) with $\xi_1=0$ depends
only on the two parameters $\xi_+$ and $\xi_-$. In the third case
($a=0$), the dilaton shift is irrelevant, so that the form field is simply
\be
F_{[n]}=b_1\mbox{vol}(\Omega_{k})\wedge
dy_1\wedge\cdots\wedge dy_{n-k}
-b_2\,\frac{dr}{r^k}\wedge dt\wedge dx_1\wedge\cdots\wedge
dx_{n-2}\,,\lb{bdF10}
\ee
which replaces (\ref{bdF1}). The three parameters are in this case
$b_1$ and $b_2$ (proportional to the magnetic and electric charges),
and the horizon radius $\xi_+$, the constant $\xi_-$ being related to
these by the condition $\e^{G_0}=1$ which reads in this case
\be\lb{cons0}
\xi_+\xi_-=\frac{(n-1)(b_1^2+b_2^2)}{4(k-1)^2}\,.
\ee

\paragraph{Second class: asymptotically LDB dyons} ($a^2=n-1$).
The solutions of this class are obtained by taking the limit $\xi_1 \to
-\infty$ in (\ref{bdm})-(\ref{bdF}). The function $f_1$ diverges in this
limit, however it enters the solution only through the combinations
$\e^{G_0}/f_1$ and $e^{a\phi_0}/f_1$ which go to the finite limits
\be
\frac{\e^{G_0}}{f_1} \to \frac{\xi}{\xi_0}\,, \quad \frac{e^{a\phi_0}}{f_1}
\to \e^{a\phi_1}\frac{\xi}{\xi_0} \,,
\ee
where we have put
\be
\xi_0 = \frac{(n-1)b_1b_2}{2(k-1)^2(\xi_+\xi_-)^{1/2}}\,, \quad
\e^{a\phi_1} = \frac{2(k-1)^2\xi_0^2}{(n-1)b_1^2}\,.
\ee

Choosing again $\xi_2 = 0$, and performing the shift on the dilaton,
together with the form rescaling
\be
\phi \to \phi - \phi_1, \quad F_{[n]} \to \e^{a\phi_1/2}F_{[n]},
\ee
the resulting solution is
\bea
ds^2&=&\left[f_{-}^2\frac{r^{k-1}}{\xi_0}\right]^{\frac1{n-1}}
\left\{-\frac{f_{+}}{f_{-}}dt^2+d{\bf x}^2\right\} \nn\\
&+&\left[f_{-}^2\frac{r^{k-1}}{\xi_0}\right]^
{-\frac1{n-1}}\bigg[(f_-^2)^{\frac1{k-1}}
\left\{\frac{dr^2}{f_{+}f_{-}}+r^2d\Omega^2_{k}\right\}+ d{\bf y}^2\bigg],
\lb{bdm2}
\\ e^{a\phi}&=&\frac{r^{k-1}}{\xi_0}\lb{bdphi2}\\
F_{[n]}&=&\sqrt{\frac2{n-1}}(k-1)\bigg[\xi_0\mbox{vol}(\Omega_{k})\wedge
dy_1\wedge\cdots\wedge dy_{n-k} \nn \\
&&\qquad\qquad-\sqrt{\xi_+\xi_-}\,\frac{dr}{r^k}\wedge
dt\wedge dx_1\wedge\cdots\wedge dx_{n-2}\bigg]\,, \lb{bdF2}
\eea
This depends on the three independent parameters $\xi_+$ and $\xi_-$ (the
locations of the outer and the inner horizons), and $\xi_0$ (the overall
scale). Again, the original parameters $b_1$ and $b_2$ have been eliminated
from the solution. The magnetic and electric charges associated with the
solution (\ref{bdm2})-(\ref{bdF2}) are
\bea
&& P = L_pL_{q-k}\Omega_k\sqrt{\frac2{n-1}}(k-1)\xi_0\,,\lb{P2}\\
&& Q = L_pL_{q-k}\Omega_k
\sqrt{\frac2{n-1}}(k-1)(\xi_+\xi_-)^{1/2}\,. \lb{Q2}
\eea
Note that the magnetic charge does not depend on the horizon radii
$\xi_+,\,\xi_-$, but only on the overall scale $\xi_0$, so that it is a
property of the linear dilaton background rather than of the black brane.
On the other hand, the electric charge does depend on these parameters, and
goes to zero in the limit $\xi_+=\xi_-=0$. The LDB metric is recovered in
this limit:
\be\lb{ldb}
ds^2=\left[\frac{r^{k-1}}{\xi_0}\right]^{\frac1{n-1}} \left[-dt^2+d{\bf
x}^2\right] + \left[\frac{r^{k-1}}{\xi_0}\right]^
{-\frac1{n-1}}[dr^2+r^2d\Omega^2_{k}+ d{\bf y}^2]\,.
\ee

The dual solutions, corresponding to the limit $\xi_2 \to +\infty$, may be
obtained from (\ref{bdm2})-(\ref{bdF2}) by the discrete S-duality:
\bea
g_{\mu\nu} \to g_{\mu\nu}, \qquad  F \to \e^{-a\phi} \ast F, \qquad  \phi
\to -\phi\,. \label{duality}
\eea
The roles of the electric and magnetic charges are then exchanged, the
electric charge being associated with the background, and the magnetic
charge with the black brane.

In both cases, we find that $\xi = \xi_-$ is generically a spacelike
singularity. However for $k = n = 2$ or 3, $\xi = \xi_-$ is a regular
inner horizon hiding a timelike singularity at $\xi = 0$.

\paragraph{Third class.}
This is obtained by taking the limit $\xi_1 = \xi_2 \to -\infty$.
In this limit the combination $\e^{G_0}/f_1f_2$ goes to the finite
limit \be \frac{\e^{G_0}}{f_1f_2} \to
\bigg(\frac{\xi}{\xi_0}\bigg)^2\,, \ee where now \be \xi_0^2 =
\frac{(n-1)b_1b_2}{2(k-1)^2}\,. \ee Taking the limit of
(\ref{bdphi}) and (\ref{eaphi0}), we obtain that the dilaton is
frozen:
\be \e^{a\phi} = \e^{a\phi_0} = \frac{b_2}{b_1}\,. \ee Thus
this third class of black dyons arises in the two cases $\phi =
\phi_0$ with either $a \neq 0$, or $a=0$. In the case $a \neq 0$,
choosing $\xi_-=0$ and performing the dilaton shift
(\ref{shift0}), we obtain the solution \bea
ds^2&=&\left[\frac{\xi}{\xi_0}\right]^{\frac2{n-1}}
\left\{-f_{+}dt^2+d{\bf x}^2\right\}
+\left[\frac{\xi}{\xi_0}\right]^ {-\frac2{n-1}}\bigg[
\frac{dr^2}{f_{+}}+r^2d\Omega^2_{k}+ d{\bf y}^2\bigg], \lb{bdm3}
\\ e^{a\phi}&=&1,\lb{bdphi3}\\
F_{[n]}&=&\sqrt{\frac2{n-1}}(k-1)\xi_0\bigg[\mbox{vol}(\Omega_{k})\wedge
dy_1\wedge\cdots\wedge dy_{n-k} \nn\\&&\qquad\qquad
-\bigg(\frac{\xi}{\xi_0}\bigg)^2\,\frac{dr}{r^k}\wedge dt\wedge
dx_1\wedge\cdots\wedge dx_{n-2}\bigg]\,, \lb{bdF3} \eea depending
on only two parameters, the horizon location $\xi_+$ and the scale
$\xi_0$. In the case $a=0$, the product $b_1b_2$ should be
replaced by $(b_1^2+b_2^2)/2$ in the definition of the parameter
$\xi_0$, while the magnetic and electric form field strengths stay
arbitrary, so that the solution given by (\ref{bdm3}),
(\ref{bdphi3}) and \bea
F_{[n]}&=&\frac{2(k-1)}{\sqrt{n-1}}\xi_0\bigg[\cos\alpha\,\mbox{vol}
(\Omega_{k})\wedge dy_1\wedge\cdots\wedge
dy_{n-k}\nn\\&&\qquad\qquad
-\sin\alpha\,\bigg(\frac{\xi}{\xi_0}\bigg)^2\,\frac{dr}{r^k}\wedge
dt\wedge dx_1\wedge\cdots\wedge dx_{n-2}\bigg]\,, \lb{bdF30} \eea
now depends on a third parameter $\alpha$. In all cases, the
electric and magnetic charges are both independent of the horizon
parameter $\xi_+$, and so are associated with the background
rather than with the black brane. The background metric  is
obtained by putting $\xi_+=0$ in (\ref{bdm3}): \be
ds^2=\left[\frac{\xi}{\xi_0}\right]^{\frac2{n-1}}
\left\{-dt^2+d{\bf x}^2\right\} +\left[\frac{\xi}{\xi_0}\right]^
{-\frac2{n-1}}\bigg[ dr^2+r^2d\Omega^2_{k}+ d{\bf y}^2\bigg]. \ee
This dyonic LDB is supported only by the antisymmetric form
electric and magnetic fluxes, the dilaton being frozen. It has
a ly simple form if $k=n$, reducing to $AdS_n\times S^n$:\be
ds^2=\left(\frac{r}{r_0}\right)^2 \left(-dt^2+d{\bf
x}_{n-2}^2\right) +\left(\frac{r_0}{r}\right)^2
dr^2+r_0^2d\Omega^2_{k}. \ee
 Returning to the spacetime (\ref{bdm3}), we see that the
two regular cases $n=2$ or 3 now correspond to geodesically
complete spacetimes, $AdS_2\times S^2$ (in a non-complete chart)
for $n=k=2$, and $BTZ \times S^3$ (with $BTZ$ the
Ba\~nados-Teitelboim-Zanelli black hole \cite{btz}) for $n=k=3$.

\section{Mass, entropy, temperature and the first law of thermodynamics}
\setcounter{equation}{0} In order to give a correct interpretation
to the black branes obtained it is useful to develop the
corresponding thermodynamics. To compute the mass in the case of
non-asymptotically flat configurations, we shall use the
quasilocal approach \cite{Brown:1992br,HaHo95,ChNe98,Booth:2000iq} as extended to
the case of the Einstein-dilaton-antisymmetric form theory in $d$
dimensions in \cite{Clement:2004ii}. The ADM decomposition \be\lb{ADM}
ds^2=-N^2dt^2+h_{ij}(dx^i+N^idt)(dx^j+N^jdt) \ee leads to a
foliation of the spacetime by spacelike $(d-1)$-surfaces
$\Sigma_t$ of metric $h_{\mu\nu}$. The surfaces $\Sigma_t$ are
themselves foliated by $(d-2)$-surfaces $\Sigma^r_t$ ($t,r$
constant) of metric $\sigma_{\mu\nu}=h_{\mu\nu}-n_\mu n_\nu$, with
$n^\mu$ the unit spacelike normal to $S^r_t$. A careful evaluation
of the field-theoretical Hamiltonian for the theory (\ref{action})
in a spacetime volume bounded by initial and final spacelike
surfaces $\Sigma_{t_i}$ and $\Sigma_{t_f}$, and a timelike surface
$r=$ constant, leads to the sum of a volume contribution which
vanishes on shell, plus a surface contribution, the quasilocal
energy, which is given in the static case ($N^i=0$) by \be\lb{qle}
E=\int_{\Sigma^r_t}\left(2\sqrt{\sigma}N\ka + (n-1)A_{ti_1\ldots
i_{n-2}} \Pi^{ri_1\ldots i_{n-2}}\right)d^{d-2}x. \ee In
(\ref{qle}), \be \ka=-\sigma^{\mu\alpha}D_{\alpha}n_\mu \ee (with
$D_{\alpha}$ the covariant derivative compatible with the metric
$h_{\mu\nu}$) is the extrinsic curvature of $\Sigma^r_t$ embedded
in $\Sigma_t$, $A_{ti_1\ldots i_{n-2}}$ are the electric
components of the $(q-1)$-potential form $A$ ($F = dA$), and \be
\Pi^{ri_1\ldots i_{n-2}}=-\sqrt{-g} \e^{a\phi}F^{tri_1\ldots
i_{n-2}} \ee are the conjugate momenta, equal to the constant
electric charges per brane volume.

The quasilocal mass may formally defined as the quasilocal energy
evaluated in the limit $r\to\infty$. However the quasilocal energy
generically diverges in this limit. This divergence may be
regularized by subtracting the contribution of a background
solution (the zero-point energy), provided  one can impose the same
Dirichlet boundary conditions on $\Sigma^r_t$ for the black solution
under consideration and for the background solution. Specifically,
this means that the boundary metric $\sigma_{ij}$ and the
non-dynamical fields (Lagrange multipliers) $N$, $N^i$ and
$A_{ti_1\ldots i_{n-2}}$ of the black solution and of the background
solution should coincide (or asymptotically coincide to a sufficient
accuracy) on the boundary \cite{HaHo95}. For the non-dynamical fields,
this requirement can be taken care of by a rescaling of time for $N$,
or a gauge transformation for $A_{ti_1\ldots i_{n-2}}$. On the other
hand, the requirement on the boundary metric strongly constrains the
choice of the background solution, which in practice must be an
extreme member of the black family of solutions. After regularization,
the quasilocal mass is
\bea\lb{qlm}
M&=&\lim_{r_b\to\infty}\int_{\Sigma^r_t}\left[2\sqrt{\sigma}N(\ka-\ka_0)
+ (n-1)A_{ti_1\ldots i_{n-2}}(\bar{\Pi}^{ri_1\ldots i_{n-2}}
-\bar{\Pi}^{ri_1\ldots i_{n-2}}_0)\right]d^{d-2}x,
\eea
where the quantities with the subscript $0$ are associated with
the background solution, and the equation of the boundary $\Sigma^r_t$
is $t =$ constant, $r=r_b$.

\paragraph{First class.}
The electric potential $A_{ti_1\ldots i_{n-2}}$ goes to zero as
$r^{1-k}$, so that the quasilocal mass is given by the purely metric
contribution (the first term in the r.h.s. of (\ref{qlm})).
The natural background for asymptotically flat black dyons is the
Minkowski spacetime, which is obtained from (\ref{bdm1}) by putting
$\xi_+ = \xi_- = 0$,
\be\lb{bg1}
ds_0^2=-dt^2+d{\bf x}^2+d\rho^2+\rho^2d\Omega^2_{k}+ d{\bf y}^2.
\ee
Note that we take care to distinguish between the generic black
radial coordinate $r$ and the background radial coordinate $\rho$.
The equation of the boundary $\Sigma^r_t$ is $r = r_b$ in
black coordinates, or $\rho = \rho_b$ in background coordinates. Because
this boundary is common, the identification of the $k$-spheres leads to
\be C_0(\rho_b) = C(r_b), \ee
which can be solved to lead to a function $\rho_b(r_b)$. The adjustment of
the ($xx$) or ($yy$) components can be simply taken care of by a
radius-dependent rescaling of the $x$ or $y$ coordinates. Then, the
computation of the extrinsic curvature leads to
\be
\ka(r_b) = -\e^{-A}\partial_r[kC + pD + (n-k)E]|_{r=r_b}
\ee
for the black metric, and
\be
\ka_0(\rho_b) = -\e^{-A_0}\partial_{\rho}[kC_0 + pD_0 +(n-k)E_0]|_{\rho=\rho_b}
\ee
for the background metric.

We obtain asymptotically, for the black metric
\bea
\ka(r_b)&\simeq&-k\xi_b^{-\frac{1}{k-1}}\left[1+\left(-\frac{\xi_+}{2}+
\frac{-4+k(n+9)+k^2(n-7)}{2k(k-1)(n-1)}\,\xi_-+\frac{3k-2}{2k(n-1)}\,
\xi_1\right)\frac1{\xi_b}\right]\,,
\eea
and for the Minkowski background
\bea
\ka_0(\rho_b)&\simeq&
-k\xi_b^{-\frac{1}{k-1}}\left[1+\left(\frac{n-k}{(k-1)(n-1)}
\,\xi_-+\frac{1}{2(n-1)}\,\xi_1\right)\frac1{\xi_b}\right]\,,
\eea
leading to the quasilocal mass
\be\lb{M1}
M = L_pL_{q-k}\Omega_k\left[k(\xi_+-\xi_-)
+\frac{2(k-1)}{n-1}(2\xi_--\xi_1)\right]\,.
\ee
This result coincides with the ADM mass.

Now we check that this value of the mass, together with the other
physical parameters of the dyonic black branes, satisfy the
generalized first law of black hole thermodynamics \cite{Rasheed:1995zv,ClLe04}
\be\lb{first}
dM=TdS+W_{h}dP+V_{h}dQ,
\ee
where $T$ and $S$ are the Hawking temperature and the black hole
entropy, $W_h$ and $V_h$ are the values of the magnetic and electric
potentials on the horizon $\xi=\xi_+$.

The entropy and the temperature are found locally in
a standard way:
\bea S&=&4\pi L_pL_{q-k}\Omega_k
\xi_+^{\frac1{n-1}}(\xi_+-\xi_-)^{\frac{k}{k-1}-\frac2{n-1}}
(\xi_+-\xi_1)^{\frac1{n-1}},\\
T&=&\frac{k-1}{4\pi}\xi_+^{-\frac1{n-1}}(\xi_+-\xi_-)^{\frac2{n-1}-
\frac1{k-1}}(\xi_+-\xi_1)^{-\frac1{n-1}}.
\eea
The electric potential and the magnetic potential (the electric
potential of the dual form) can be written out from the form field
(\ref{bdF1}) or (\ref{bdF10}) as follows
\bea
W=\sqrt{\frac2{n-1}}\frac{\sqrt{(\xi_+-\xi_1)(\xi_--\xi_1)}}{\xi-\xi_1}\,,
\quad V=\sqrt{\frac2{n-1}}\frac{\sqrt{\xi_+\xi_-}}{\xi}\,,
\eea
for $a^2=n-1$, and for arbitrary $a\neq0$ with $\xi_1=0$, or
\bea
W=\frac1{k-1}\frac{b_1}{\xi}\,, \quad
V=\frac1{k-1}\frac{b_2}{\xi}\,,
\eea
for $a=0$ (implying $\xi_1=0$). In all cases we find that the
generalized first law (\ref{first})
is satisfied under independent variation of the parameters
$\xi_+$, $\xi_-$ and $\xi_1$ (case $a^2=n-1$); $\xi_+$ and $\xi_-$
(case $a\neq0$ with $\xi_1=0$);  or $\xi_+$, $b_1$ and $b_2$
(case $a=0$, where the relation (\ref{cons0}) between the parameters
should be taken into account).

\paragraph{Second class.} Again the electric potential of the electric
black dyon (\ref{bdF2}) goes to zero as
$r^{1-k}$, so that the quasilocal mass is given by the purely metric
contribution. The natural background is the magnetically charged LDB
$\xi_+=\xi_-=0$ (\ref{ldb}). We obtain asymptotically, for the
extrinsic curvature of the black metric
\be
\ka(r_b) \simeq -\frac{1+k(n-2)}{n-1}\bigg(\frac{\xi_b}{\xi_0}\bigg)^
{\frac{1}{2(n-1)}}\xi_b^{-\frac{1}{k-1}}\left[1-\left(\frac{\xi_+}{2}-
\frac{n+1+k(n-3)}{2(k-1)(n-1)}\xi_-+\frac{k-1}{1+k(n-2)}\xi_-\right)
\frac1{\xi_b}\right]\,,
\ee
and for that of the linear dilaton background
\be
\ka_0(\rho_b)\simeq-\frac{1+k(n-2)}{n-1}\bigg(\frac{\xi_b}{\xi_0}\bigg)^
{\frac{1}{2(n-1)}}\xi_b^{-\frac{1}{k-1}}
\left(1+\frac{n-k}{(k-1)(n-1)}\frac{\xi_-}{\xi_b}\right)\,,
\ee
leading to the quasilocal mass
\be\lb{M2}
M=L_pL_{q-k}\Omega_k\left[\left(k-\frac{k-1}{n-1}\right)
(\xi_+-\xi_-)+2\frac{k-1}{n-1}\xi_-\right]\,.
\ee

Bearing in mind that the magnetic charge is a property of the linear
dilaton background and thus should not be varied, the first law for
asymptotically LDB black dyons reads
\be\lb{firste}
dM=TdS+V_{h}dQ,
\ee
The entropy and temperature are
\bea
&& S =4\pi L_pL_{q-k}\Omega_k(\xi_0\xi_+)^{\frac1{n-1}}
(\xi_+
-\xi_-)^{\frac k{k-1}-\frac2{n-1}}, \lb{S2}\\
&&T=\frac{k-1}{4\pi}(\xi_0\xi_+)^{-\frac1{n-1}}(\xi_+-\xi_-)^
{\frac2{n-1}-\frac1{k-1}}\,, \lb{T2}
\eea
the electric potential is
\be
V=\frac1{k-1}\frac{b_2}{\xi}\,, \lb{V2}
\ee
and the electric charge is given by (\ref{Q2}). These quantities satisfy
the first law (\ref{firste}) under independent variation of the
parameters $\xi_+$ and $\xi_-$, the scale parameter $\xi_0$,
associated with the linear dilaton background, being held fixed.

In the case of the dual magnetic black dyon, the constant electric
field $\Pi^{ri_1\ldots i_{n-2}}$ is identical to that of the electric
linear dilaton background, so that the electric contribution to the
quasilocal mass (\ref{qlm}) is identically zero, and the quasilocal
mass is again given by (\ref{M2}). The first law appropriate for this case,
\be\lb{firstm}
dM=TdS+W_{h}dP,
\ee
is again satisfied (with the magnetic potential $W$ and the magnetic
charge $P$ given by (\ref{V2}) and (\ref{Q2})) provided the scale
parameter $\xi_0$, proportional to the electric charge of the linear
dilaton background, is held fixed.

\paragraph{Third class.} In this case, the constant electric field is
again identical to that of the background $\xi_+ = 0$, so that the
quasilocal mass is given by the sole metric contribution. From the
asymptotic extrinsic curvature of the black metric
\be
\ka(r_b) \simeq -\frac{2+k(n-3)}{n-1}\bigg(\frac{\xi_b}{\xi_0}\bigg)^
{\frac{1}{n-1}}\xi_b^{-\frac{1}{k-1}}\left[1-\frac{\xi_+}{2\xi_b}\right]\,,
\ee
we obtain (in this case $\rho_b = r_b$)
\be\lb{M3}
M=L_pL_{q-k}\Omega_k\left(k-2\frac{k-1}{n-1}\right)
\xi_+\,.
\ee
The mass is zero for $n=k=2$ (the solution in this case is
Bertotti-Robinson), and positive in
the other cases ($n>2$ and $k \ge 2$, with $n \ge k$).

The entropy and temperature are
\bea
&& S =4\pi L_pL_{q-k}\Omega_k\bigg(\frac{\xi_+}{\xi_0}\bigg)^{-\frac2{n-1}}
\xi_+^{\frac{k}{k-1}}\,,\lb{S3}\\
&&T=\frac{k-1}{4\pi}\bigg(\frac{\xi_+}{\xi_0}\bigg)^{\frac2{n-1}}
\xi_+^{-\frac1{k-1}}\,, \lb{T3}
\eea
and the first law takes the simple form
\be
dM=TdS,
\ee
since both charges belong to the background, and so should not be
varied.

\section{Conclusions}
In this paper we have constructed new non-asymptotically flat
$p$-brane solutions which possess a regular event horizon and
which approach the linear dilaton
background at spatial infinity. The latter is a supersymmetric
solution of the supergravity equations with a non-zero flux of
the  antisymmetric form field. More precisely, we have shown that
there exist magnetically charged $p$-branes on an electric LDB,
electrically charged branes on a magnetic LDB and uncharged branes
on a LDB with both electric and magnetic fluxes. Together with the
usual asymptotically flat dyonic branes, these configurations
exhaust all possibilities for brane solutions free of naked
singularities and involving both electric and magnetic sectors of the
(unique) form field.

The physical interpretation of non-asymptotically flat dyonic
solutions has some subtleties related to the nature of the charge
parameters. It turns out that at least one of the two charge
parameters must be attributed to the background, not to the brane
itself. This is clearly seen from the first law of thermodynamics,
which is derived using the generalized formalism of quasilocal
charges.

Black brane solutions on a fluxed linear dilaton background
describe the thermal phase of the QFT involved into the
corresponding DW/QFT correspondence \cite{Be99}. Previously
\cite{Clement:2004ii} we have found configurations of this kind
involving neutral branes on a purely magnetic or purely electric
LDB. Now we see that there exist also charged branes on a LDB
with a dual flux (i.e. electric branes on a magnetic LDB and vice
versa) as well as uncharged branes on a dyonic LDB. Their role
in the DW/QFT correspondence requires further study.

{\bf Acknowledgments.}  D.G. is grateful to LAPTH Annecy for
hospitality in June 2005 while the paper was finalized. D.G. also
thanks J.M. Nester and C.M. Chen for hospitality and useful discussions
during his visit to NCU, Taiwan. The work
of D.G. and D.O. was supported in part by the RFBR grant
02-04-16949.

\end{document}